# Security Metrics in Industrial Control Systems

Zachary A. Collier[1], Mahesh Panwar[2], Alexander A. Ganin[3], Alex Kott[4], Igor Linkov[1*]

[1] US Army Engineer Research & Development Center, Concord, MA, USA

[2] Contractor to US Army Engineer Research & Development Center, Concord, MA, USA

[3] University of Virginia, Charlottesville, VA, USA

[4] US Army Research Laboratory, Adelphi, MD, USA

*Corresponding Author, Igor.Linkov@usace.army.mil

## 1.1 Introduction

Risk – the topic of the previous chapter – is the best known and perhaps the best studied example within a much broader class of cyber security metrics. However, risk is not the only possible cyber security metric. Other metrics such as resilience can exist and could be potentially very valuable to defenders of ICS systems.

Often, metrics are defined as measurable properties of a system that quantify the degree to which objectives of the system are achieved. Metrics can provide cyber defenders of an ICS with critical insights regarding the system. Metrics are generally acquired by analyzing relevant attributes of that system.

In terms of cyber security metrics, ICSs tend to have unique features: in many cases, these systems are older technologies that were designed for functionality rather than security. They are also extremely diverse systems that have different requirements and objectives. Therefore, metrics for ICSs must be tailored to a diverse group of systems with many features and perform many different functions.

In this chapter, we first outline the general theory of performance metrics, and highlight examples from the cyber security domain and ICS in particular. We then focus on a particular example of a class of metrics that is different from the one we have considered in earlier chapters. Instead of risk, here we consider metrics of resilience. Resilience is defined by the National Academy of Sciences

(2012) as *"The ability to prepare and plan for, absorb, recover from, or more successfully adapt to actual or potential adverse events".*

This chapter presents two approaches for the generation of metrics based on the concept of resilience using a matrix-based approach and a network-based approach. Finally, a discussion of the benefits and drawbacks of different methods is presented along with a process and tips intended to aid in devising effective metrics.

## 1.2 Motivation

Under President George W. Bush, the Department of Energy issued best practices for improved industrial control system (ICS) security (US Department of Energy, 2002). Some of these include taking steps such as "disconnect unnecessary connections to the SCADA network", "establish a rigorous, ongoing risk management process" and "clearly identify cyber security requirements." Additionally, Executive Order 13636, signed by President Barack Obama in 2013, brought forth the issue of cyber security and resilience, and proposed the development of a risk-based "Cybersecurity Framework" (EO 13636, 2013).

The framework was presented by the National Institute of Standards and Technology (NIST) and offers organizations guidance on implementing cybersecurity measures.

Despite existing guidelines and frameworks, designing and managing for security in cyber-enabled systems remains difficult. This is in large part due to the challenges associated with the *measurement* of security. Pfleeger and Cunningham (2010) outline nine reasons why measuring security is a difficult task as it relates to cybersecurity in general, but all of which also apply to the security of ICS domain (Table 1).

Pfleeger and Cunningham (2010) note that one way to overcome these challenges is to thoughtfully develop a clear set of security metrics. Unfortunately, this lack of metrics happens to be one of the greatest barriers to success in implementing ICS security. When ICSs were first implemented, "network security was hardly even a concern" (Igure et al, 2006). Although efforts are being made to draft and enact cyber security measures, that gap has yet to be closed, even at a time of greater risk.

Table 1: Challenges with Cybersecurity Measurement (*adapted from Pfleeger & Cunningham, 2010*)

| Challenge | Description |
| --- | --- |
| We can't test all security requirements | It is not possible to know all possible configurations and states of the system, intended uses and unintended misuses from users, etc. |
| Environment, abstraction, and context affect security | Systems are built to evolve as they process new information, and not all system changes are derived from malicious sources |
| Measurement and security interact | Knowledge about a system's vulnerabilities and safeguards can affect the types of further security measures implemented, as well as modify the risks that users are willing to take |
| No system stands alone | Systems are networked to interact with other cyber systems and assets |
| Security is multidimensional, emergent, and irreducible | Security exists at multiple levels of system abstraction, and the security of the whole system cannot be determined from the security of the sum of its parts |
| The adversary changes the environment | Developing an accurate threat landscape is difficult due to adaptive adversaries who continually develop novel attacks |
| Measurement is both an expectation and an organizational objective | Different organizations with different missions and preferences place differing values on the benefits of security |
| We're overoptimistic | Users tend to underestimate the likelihood that their system could be the target of attack |
| We perceive gains differently than losses | Biases in interpreting expected gains and losses based on problem framing tend to affect risk tolerance and decision making under uncertainty in predictable but irrational ways |

## 1.3 Background on Resilience Metrics

### 1.3.1 What Makes a Good Metric?

According to the management adage, "what gets measured gets done". As such, well-developed metrics can assist an organization in reaching its strategic goals (Marr, 2010). Reichert et al. (2007) define metrics as "measurable properties that quantify the degree to which objectives have been achieved". Metrics provide vital information pertaining to a given system, and are generally acquired by way of analyzing relevant attributes of that system. Some researchers and practitioners make a distinction between a measure and a metric (Black et al., 2008, Linkov et al., 2013a), whereas others may refer to them as performance measures (Neely et al., 1997), key performance indicators (Marr, 2010) or strategic measures (Allen, 2011). For the purposes of this chapter, these are referred to generally as metrics.

When used efficiently, metrics can help to clarify one's understanding of the processes of a particular area of a system, and from there, provide information for external review and assist towards further improvement, among other outputs (Marr, 2010). This can be done by establishing benchmarks for a given metric, where thresholds or ranges can be established (Black et al., 2008). Benchmarks, or standards, help form the basis for decision making and taking corrective action (Williamson, 2006).

A critical element in eliciting a meaningful metric is to gather the relevant information about one's system and to align that metric with measurable goals and strategic objectives which lie within the scope of a given project or the domain of a particular organizational structure (Beasley et al. 2010, Neely et al. 1997). There is also the issue of scale and adaptability. Smaller organization may have metrics dealing with rudimentary security measures, but as they grow larger, these measures may need to be scaled appropriately to deal with the security needed for a larger organization (Black et al., 2008).

There are key elements that contribute to producing a successful metric. Metrics should be actionable: they are not simply about measuring numerous attributes of a project; merely gathering information without a goal in mind will not provide a discernible solution (Marr, 2010). Such information in and of itself would not be substantial enough to be considered a metric. Gathering relevant metrics requires delving deeper into the issues faced by a given system and asking pertinent questions which can lead to actionable improvement. These include questions such as "Does it link to strategy? Can it be quantified? Does it drive the right behavior?" (Eckerson, 2009). From these, one can obtain metrics which can in turn inform actionable results. Table 2 summarizes the desirable characteristics of metrics in general terms, and apply to all types of systems including ICSs.

Table 2: Characteristics of Good Metrics (*adapted from McKay et al. 2012, Keeney and Gregory 2005*)

| Characteristic | Description |
| --- | --- |
| Relevant | Metrics are directly linked to decision making goals and objectives |
| Unambiguous | Consequences of alternatives can be clearly measured by metrics |
| Direct | Metrics clearly address and describe consequences of interest |
| Operational | Data exist and are available for the metric of interest |
| Understandable | Metrics can be understood and communicated easily |
| Comprehensive | The set of metrics address a complete suite of goals and consequences |

Metrics may be described as natural, constructed, or proxy. Natural metrics directly describe an objective in units that are straightforward (e.g., dollars as a metric for "costs associated with ICS downtime"). Constructed metrics may be used when natural metrics do not exist (e.g., scales from 1 to 10 where each number corresponds to a defined level of ICS performance), and usually incorporate expert judgment. Proxy metrics can be used to indirectly measure an objective (e.g., the number of users with certain administrative privileges as a proxy for access) (McKay et al. 2012, Keeney and Gregory 2005).

There are different types of information that metrics gauge and the project team has the responsibility of appropriately selecting and evaluating them. These can be separated into quantitative, semi-quantitative and qualitative approaches. Quantitative metrics have measurable, numerical values attached to them. Semi-quantitative metrics are not strictly quantifiable but can be categorized. Qualitative metrics provide non-numeric information, for example in the form of aesthetics.

### 1.3.2 Metrics for IT Systems

As described above in Table 1, cyber systems provide unique challenges. In particular, the cyber domain extends beyond just the immediate system and requires a holistic viewpoint, with many different technical and human factors to be accounted for (Collier et al., 2014). Threats to the system are also constantly evolving and growing in sophistication, and as a result, there is a high degree of adaptability required in order to remain current. Due to the constantly evolving threat space, there is often little historical data for potential threats (Collier et al., 2014).

With cyber metrics, a significant number of the main issues are tailored towards security and resilience. The Defense Science Board (2013) argues that effective cyber metrics should be broad enough to fit different types of systems, yet also be precise enough to dial down into the specifics of a given system. The following are some examples of cybersecurity metrics currently in use.

The Common Vulnerability Scoring System (CVSS) was introduced to provide various organizations with actionable information in regards to assessing IT vulnerabilities (Mell et al., 2007). CVSS groups their metrics into three categories, namely Base, Temporal, and Environmental metrics.  A few of these security metrics include Collateral Damage Potential, Target Distribution, Report Confidence, Exploitability, Access Complexity, Access Vector, Authentication, Integrity Impact, Availability Impact, and Confidentiality Impact (Mell et al., 2007).  There are general scoring tips for the way that vulnerabilities are assessed; vulnerabilities are not scored based on interactions with other vulnerabilities, rather, they are scored independently. The main measure of vulnerability is its impact on the key service. Vulnerabilities are scored according to commonly used privileges, which might be a default setting in certain situations. If a vulnerability can be exploited by multiple exploits, it is scored with the exploit that will present the maximum impact (Mell, et al., 2007). CVSS allows vulnerability scores to be standardized, and Base metrics are normalized on a scale of 0 – 10. They can be optionally refined by including values from Temporal and Environmental metrics.

The Center for Internet Security (CIS) has also established metrics for organizations to use (CIS, 2010). CIS has divided their metrics into six critical business functions. These are Incident Management, Vulnerability Management, Patch Management, Configuration Management, Change Management and Application Security. It also recognizes hierarchies and interdependencies of metrics, for instance citing management metrics as being of primary importance to an organization, while noting that some of those metrics may depend on the prior implementation of technical metrics (CIS, 2010). Some of the metrics include Cost of Incidents and Patch Policy Compliance. Cost of Incidents refers to a number of potential losses, such as customer lists or trade secrets under a "direct loss" and a "cost of restitution", for example in the event that fines are levied due to an incident.  This is measured by the summation of the numerical values of all the costs associated with the metric. Examples relating to security include Mean Time to Incident Discovery, Mean Time Between Security Incidents and Mean Time to Incident Recovery (CIS, 2010). For an example of measurement, Mean Time to Incident Discovery measures the summation of the time between incidents and discoveries of incidents, divided by total number of incidents recovered during those time frames (CIS, 2010).

The Cybersecurity Framework developed by NIST stemming from EO 13636 was released in February 2014 (NIST 2014a). The final Cybersecurity Framework consists of a Framework Core, which presents a set of five "concurrent and continuous Functions – Identify, Protect, Detect, Respond, Recover" (NIST 2014). These functions are the "high-level, strategic view of the lifecycle of an organization's management of cybersecurity risk," which feature subsequent categories and subcategories for the functions, relating to outcomes and activities (NIST 2014). For example, the Respond function consists of five categories, among which includes Mitigation. Mitigation is then further subdivided into metrics related to containing and eradicating incidents. The Framework Core is used as a scorecard of progress – the current guidance calls for first developing an organization's Current Profile, which consists of assigned scores based on the organization's performance in each of the categories and subcategories. This Current Profile is then compared to a Target Profile, representing the desired state of the organization in each of the same categories and subcategories. The shortfalls between these profiles can be viewed as gaps in an organization's cyber-risk management capabilities which can inform prioritization of corrective measures (NIST 2014; Collier et al. 2014).

The Software Engineering Institute (SEI) at Carnegie Mellon University developed a framework for assessing operational resilience which features a set of Top Ten Strategic Measures, which aim to be mapped down to the level of specific Process Area measures (Allen and Curtis, 2011). Under the heading of High-Value Services and Assets, one of the measures is related to the percentage of high-value services that do not satisfy their assigned resilience requirements (Allen and Curtis, 2011). The SEI framework also contains a large amount of resilience measures, spanning 26 different Process Areas. For example, under the Process Area of Environmental Control, there are measures such as Percentage of Facility Assets that have been Inventoried, Elapsed Time Since the Facility Asset Inventory was Reviewed, and Elapsed Time Since Risk Assessment of Facility Assets Performed (Allen and Curtis, 2011), where the term "assets" applies to high-value services. These are presented in a table with traceability, assigning an identification number to each metric along with their applicability to goals within the Process Areas.

MITRE proposed a framework entitled Cyber Resiliency Engineering Framework, which, among its goals aims to "motivate and characterize cyber resiliency metrics" (Bodeau 2011). The framework contains four Cyber Resiliency Goals: Anticipate, Withstand, Recover, and Evolve. There are a total of eight objectives which are a subset of the goals. For example Anticipate has three objectives: Predict, Prevent, and Prepare (Bodeau, 2011). This hierarchy can be used to inform and categorize the appropriate resilience metrics. These are meant to be performed simultaneously, and bear a resemblance to the NIST framework mentioned earlier.

### 1.3.3 Metrics for ICS Networks

The above metrics were developed for "cyber" systems generally speaking, not specifically for ICSs, although they can be tailored with ICSs in mind. ICSs in particular are a unique case; in many situations, these systems have older models, and were designed for functionality rather than security (US Department of Energy, 2002). They constitute a diverse group of systems that have different requirements for their various operations (Pollet, 2002).

Specifically as it relates to ICSs, time, safety and continuation of services are of great importance, since many systems are in a position where a failure can result in a threat to human lives, environmental safety, or production output (Stouffer, 2011). Since these risks are different than those faced by information technology (IT) systems, different priorities are also necessary. Examples of some unique considerations in comparison to cyber security include the longer lifespan of system components, physically difficult to reach components, and continuous availability requirements (Stouffer, 2011). Additionally, these systems typically operate in separate fields than cybersecurity, such as in the gas and electric industries, and so metrics must be adapted to fit these different organizational structures (McIntyre et al., 2007). Critical infrastructures are common for ICSs, and as a result "downtime and halting of production are considered unacceptable" (McIntyre et al., 2007).

Stouffer et al. (2011) compare the differences between information technology (IT) system and ICSs, focusing on the safety-critical nature of many ICS networks. For example, "high delay and jitter may be acceptable" as a performance requirement for IT systems, whereas for ICSs, it may not be acceptable (Stouffer, 2011). This is due to the fact that there is a time-critical nature to ICSs, whereas for IT systems there is high throughput, allowing for some jitter (Stouffer, 2011).  Similarly, for IT, "systems are designed for use with typical operating systems" and for ICSs, there are "differing and possibly proprietary operating systems, often without security capabilities built in". There are also availability requirements, in that sometimes an IT strategy may require restarting or rebooting a process, something which, for ICS processes, requires more careful planning as unexpected outages and quickly stopping and starting a system are not acceptable solutions (Stouffer, 2011). With these key differences between the two domains, there are varying levels of adaptation needed in order to begin the process of securing ICS networks.

The US National Security Agency (NSA) drafted a framework for ICS networks, focusing on potential impact and loss relating to a network compromise (NSA, 2010). They suggested assigning loss metrics incorporating NIST's framework:  compromises pertaining to Confidentiality, Integrity and Availability for each network asset (NSA, 2010). A Confidentiality compromise is defined as an "unauthorized release or theft of sensitive information" e.g. theft of passwords (NSA, 2010). An Integrity compromise is defined as an "unauthorized alteration or manipulation of data", e.g. manipulation of billing data (NSA, 2010). An Availability compromise is defined as a "loss of access to the primary mission of a networked asset" e.g. deletion of important data from a database (NSA, 2010). These may also be streamlined into one metric, using the highest value (e.g. of Low, Moderate or High) among the three areas.

The assignment of a threat metrics at each potential attack vector was suggested, but specific examples were not provided. Five threat sources were identified: Insiders, Terrorists or Activists, Hackers or Cyber-Criminals, Nation/State Sponsored Cyber-Warfare and Competitors (NSA, 2010). Both loss and threat metrics can be rated on a constructed scale (Low, Moderate or High) and given a numeric rating on a set scale. It was mentioned that the important consideration is to have a scale, and that the number of graduations in the scale is not important, so long as the constructed scale remain consistent (e.g. a potential for loss of life will rank as High) (NSA, 2010). Combining results of metrics was also discussed as a possibility. As an example, for a given point in the network, a Loss Metric is assigned a score of High on the constructed scale (3) and a Threat metric at that same network point is rated at Moderate (2). From this, one can arrive at a composite priority value, which is simply the sum of those two scores. Other such points can be evaluated and then prioritized and ranked (NSA, 2010). The scoring methodology is a basic example, (and not the only method - weighing metrics was listed as a possibility (NSA, 2010)) and more robust methods can be devised.

Boyer and McQueen (2008) devised a set of ideal-based technical metrics for control systems. They examined seven security dimensions and present an ideal, or best case scenario, for each of them. The ideals are Security Group Knowledge, Attack Group Knowledge, Access, Vulnerabilities, Damage Potential, Detection, and Recovery. For the Access dimension, the ideal states that the system is inaccessible to attack groups. The security dimension of Vulnerabilities has an ideal stating that the system has no vulnerabilities (Boyer and McQueen, 2008). By the very nature of an ideal, these may be impossible to achieve and maintain in the real world. But from them, metrics were devised that could best represent the realization of these ideals. Under the vulnerability dimension, the metric Vulnerability Exposure is defined as "the sum of known and unpatched vulnerabilities, each multiplied by their exposure time interval." It was suggested that this metric could be broken down into separate metrics for different vulnerability categories, as well as including a prioritization of vulnerabilities, citing CVSS. Under the Access dimension, there is the metric Root Privilege Count, which is the count of all personnel with key privileges, arguing in favor of the principle of least privilege, which states that "every program and every privileged user of the system should operate using the least amount of privilege necessary to complete the job" (Saltzer, 1974). This logical ordering of metrics within the scope of ideals can be of value to those wishing to devise their own set of metrics.

The ideal-based metrics (Boyer and McQueen, 2008) also acknowledge the physical space of ICS networks. The metric Rogue Change Days, which is the number of changes to the system multiplied by the number of days undetected, includes Programmable Logic Controllers and Human-Machine Interfaces and other ICS related systems. Component Test Count, a metric measuring the number of control system components which have not been tested is a simple measure, but of significance due to numerous components in use in an ICS system.

Within the ideals, the metric of Attack Surface (defined by Manadhata and Wing (2011) as "the subset of the system's resources (methods, channels, and data) potentially used in attacks on the system") was determined to not be developed enough for real world use. Boyer and McQueen further argue that "a credible quantitative measure of security risk is not currently feasible" (Boyer and McQueen 2008). But with the inclusion of a theoretical metric, and a framework for security, this demonstrates a forward thinking attitude that can be built upon by those aiming to establish their own security protocols. This represents important future work for the ICS and security communities. Comparisons between the NSA approach and the approach outlined by Boyer and McQueen are presented in Table 3.

Table 3: Comparison between ICS Metrics

|  | **National Security Agency (2010)** | **Boyer and McQueen (2008)** |
|---|---|---|
| **Focus** | Loss and Threat focused Metrics (p. 10, 15) | Quantitative technical metrics (p.1), ideal based: attempted to have metrics that could strive toward ideal scenarios within seven security areas |
| **Amount** | Three loss metrics (per networked asset), one Threat metric (per potential attack vector) | 13 total metrics (suggested total: less than 20) |
| **Applied or Theoretical** | Suggests deployable metrics | Discusses both deployable and theoretical metrics (p. 10, 11) |
| **Quantitative or Qualitative** | Semi-qualitative (suggests High, Medium, Low, with allowance for numeric attachment to these values) | Does not focus on qualitative metrics (p. 1), but on quantitative metrics |
| **Combination of Metrics** | Presents method to combine results of metric scores for ranking | No combination of metrics |
| **Consequence Considerations** | Loss Metrics are related to Confidentiality, Integrity, Availability | Acknowledges the purpose of security is protection of Confidentiality, Integrity and Availability (p. 4) |

Complementary research to metrics development in the ICS realm is currently being conducted. One such effort is to develop a standardized taxonomy of cyber attacks on SCADA systems (Zhu et al., 2011). A common language for describing attacks across systems can facilitate the development of further threat and vulnerability metrics for ICSs. In addition, the development of a national testbed for SCADA systems is being developed by the Department of Energy which will enable the modeling and simulation of various threat and vulnerability scenarios, which will allow researchers to develop a better understanding of what metrics may or may not be useful in monitoring and management of these systems (US Department of Energy, 2009). Another development related to metrics research is the investigation of tradeoffs between certain critical metrics. One example is between optimizing system performance with system security, where additional security measures may result in reduced performance. Zeng & Chow (2012), developed an algorithmic technique to determine the optimal tradeoff between these two metrics, and the method can be extended to tradeoffs between other metrics as well.

## 1.4 Approaches for ICS Metrics

While various frameworks and sets of metrics exist, such as the ones mentioned in the previous section, it can be difficult for managers and system operators to decide whether to adopt or modify an existing set, or to create an entirely new set of metrics. Balancing the tradeoffs between generalizable metrics and specific system-level and component-level metrics can be challenging (Defense Science Board, 2013). The following approaches provide a structured way to think about developing metrics, allowing users to leverage existing metrics but also identify gaps where new metrics may need to be created. The use of such structured and formalized processes requires the thoughtful analysis of the systems being measured, but also how they relate to the broader organizational context, such as goals, constraints, and decisions (Marr, 2010). Moreover, the development of a standardized list of questions or topics helps to simplify the process of designing a metric. The development of metrics should be a smooth process, and such a list can provide insight into the "behavioral implications" of the given metrics (Neely et al. 1997).

### 1.4.1 Cyber Resilience Matrix Example

The first method is based on the work of Linkov et al. (2013a). Unlike traditional risk-based approaches, this approach takes a resilience-centric theme. Much has been written elsewhere on the relative merits of a resilience-focused approach (see Linkov et al., 2013b, 2014; Collier et al. 2014; Roege et al. 2014; DiMase et al. 2015), but we shall briefly summarize the argument here. Traditional risk assessment based on the triplet formulation proposed by Kaplan and Garrick (1981) becomes difficult to implement in the cybersecurity context due to the inability to frame and evaluate multiple dynamic threat scenarios, quantify vulnerability against adaptive adversaries, and estimate the long-term and widely distributed consequences of a successful attack. Instead of merely hardening the system against potential known threats in a risk-based approach, the system can be managed from the perspective of resilience, which includes the ability of one or more critical system functionalities to quickly "bounce back" to acceptable levels of performance. As a result, a resilient system can withstand and recover from a wide array of known and unknown threats through processes of feedback, adaptation, and learning.

Following this thought process, Linkov et al. (2013a) established a matrix-based method. On one axis, the steps of the event management cycle identified as necessary for resilience by the National Academy of Sciences (2012) are listed, and include Plan/Prepare, Absorb, Recover, and Adapt. Note that the ability to plan/prepare is relevant before an adverse event, and the other capabilities are relevant after disruption. On the other axis are listed the four domains in which complex systems exist as identified by Alberts (2002), and include Physical, Information, Cognitive, and Social domains. The Physical domain refers to the physical resources and capabilities of the system. The Information domain refers to the information and data that characterize the Physical domain. The Cognitive domain describes the use of the other domains for decision making. Finally, the Social domain refers to the organizational structure and communication systems for transmitting information and making decisions (Alberts 2002).

Together, these axes form a set of cells that identify areas where actions can be taken in specific domains to enhance the system's overall ability to plan for, and absorb, recover, and adapt to, various threats or disruptions (Figure 1). Each cell is designed to answer the question: "How is the system's ability to [plan/prepare for, absorb, recover from, adapt to] a cyber disruption implemented in the [physical, information, cognitive, social] domain?" (Linkov et al. 2013a).

|  | Plan & Prepare | Absorb | Recover | Adapt |
|---|---|---|---|---|
| **Physical** |  |  |  |  |
| **Information** |  |  |  |  |
| **Cognitive** |  |  |  |  |
| **Social** |  |  |  |  |

Figure 1: Generic Resilience Matrix

A resulting set of 49 metrics are produced that span the various cells of the matrix, and selected metrics are shown in Table 4 (see Linkov et al. 2013a for the complete list). Metrics are drawn from several sources and are meant to be general and not necessarily comprehensive. For example, under Adapt and Information, a metric states "document time between problem and discovery, discovery and recovery," which has a parallel to the Mean Time to Incident Discovery within SEI's guidance. The metrics under Plan and Information, related to identifying internal and external system dependencies can be compared to the Temporal Metric of Access Complexity from CVSS, which relates to how easily a vulnerability can be exploited. The metric under Prepare and Social presents a simple yet important message that holds true in all of the frameworks: "establish a cyber-aware culture."

The resilience matrix approach described in Linkov et al. (2013a) has several strengths in that the method is relatively simple to use and once metrics have been generated, it can serve as a platform for a multi-criteria decision aid (Collier & Linkov, 2014). It has the potential to serve as a scorecard in order to capture qualitative information about a system's resilience, and aid managers and technical experts in identifying gaps in the system's security. However, the resilience matrix does not capture the explicit temporal nature of resilience (i.e., mapping the critical functionality over time) or explicitly model the system itself. In this regard, it can be viewed as a high level management tool that can be used to identify a snapshot where more detailed analyses and modeling could potentially be carried out.

Table 4: Selected Cybersecurity Metrics Derived from the Resilience Matrix (adapted from Linkov et al., 2013a).

|  | Plan/Prepare | Absorb | Recover | Adapt |
| --- | --- | --- | --- | --- |
| Physical | Implement controls/sensors for critical assets and services | Use redundant assets to continue service | Investigate and repair malfunctioning controls or sensors | Review asset and service configuration in response to recent event |
| Information | Prepare plans for storage and containment of classified or sensitive information | Effectively and efficiently transmit relevant data to responsible stakeholders/decision makers | Review and compare systems before and after the event | Document time between problem and discovery, discovery and recovery |
| Cognitive | Understand performance trade-offs of organizational goals | Focus effort on identified critical assets and services | Establish decision making protocols or aids to select recovery options | Review management response and decision making processes |
| Social | Establish a cyber-aware culture | Locate and contact identified experts and responsible personnel | Determine liability for the organization | Evaluate employees response to event in order to determine preparedness and communications effectiveness |

### 1.4.2 Network Simulation Example

The second method is based on modeling of complex cyber and other systems as interconnected networks, where a failure in one sector can cascade to other dependent networks and assets (Vespignani, 2010). This is a reasonable assumption for ICS networks; for example, a disruption of the electrical grid can directly impact dependent sectors such as the network controlling ICS devices leading to a cascade of failures as it is believed to have happened during the Italian blackout in 2003 (Buldyrev et al., 2010). Thus the assessment of the security of a single ICS network should be viewed in the context of a larger network of interdependent systems.

Ganin et al. (2015) took this network-oriented view in developing a methodology to quantitatively assess the resilience (and thus security) of networked cyber systems. They built upon the National Academy of Sciences (2012) definition of resilience as a system property that is inherently tied to its ability to plan for, absorb, recover from, and adapt to adverse events. In order to capture the state of the system the authors propose to use the concept of critical functionality defined as a time-specific performance function of the system considered and derived based on the stakeholder's input. For instance in the network of power plants, the critical functionality might represent the total operational capacity. In the network of computers it might represent the fraction of servers and services available. Values of critical functionality are real numbers from 0 to 1. Other key elements to quantify resilience are the networked system's topology and dynamics; the range of possible adverse events (for example, a certain damage to nodes of the network); and the control time $T_C$ (that is the time range over which the performance of the system is evaluated). Then the dependency of the critical functionality (averaged over all adverse events) over time is built. Ganin et al. (2015) refer to this dependency as the resilience profile. As it is typically computationally prohibitive or not possible at all (in case of continuous variables defining nodes' states) to consider all the ways an adverse event can happen, it is suggested to utilize a simulation based approach with Monte-Carlo sampling.

Given its profile in normalized time (where time $T_C$ is taken to be 1), the resilience of the network can be measured as the area under the curve (yellow region in Figure 2). This allows mapping of the resilience to real values ranging between 0 and 1.

Another important property of the system is obtained by finding the minimum of the average critical functionality. Some researchers refer to this value as robustness $M$ (Cimellaro et al., 2010), while Linkov et al. (2014) note that $1 - M$ corresponds to the measure of risk.

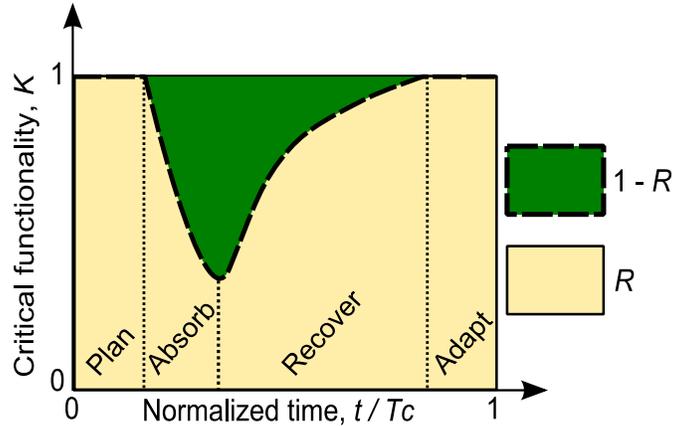

Figure 2: A generalized resilience profile, where a system's resilience is equal to the area below the critical functionality curve (adapted from Ganin et al., 2015).

In their paper Ganin et al. (2015) illustrated the approach on a directed acyclic graph. Each level in this graph represents a set of nodes from certain infrastructure system (e.g. electrical grid, computers etc.). Nodes of different levels are connected by directed links representing a dependency of the destination node on the source node. In the simplest case a node in a certain level requires supply (or a dependency link) from a node in each of the upper levels and does not depend on any nodes in the lower levels. Other parameters of the model include node recovery time ($T_R$) – a measure of how quickly a node can return to an active state after it's been inactivated as a result of an adverse event; redundancy ($p_m$) – the probability controlling the number of additional potential supply links from upper levels to lower levels; and switching probability ($p_s$), controlling ease of replacement of a disrupted supply link with a potential supply link. These parameters could be extended to other situations to inform how a system may display resilient behavior, and thus increasing the security of the system as a whole.

The authors found that there is strong synergy between $p_m$ and $p_s$; increasing both factors together produces a rapid increase in resilience, but increasing only one or the other variable will cause the resilience metric to plateau. Resilience is strongly affected by the temporal switching time factor, $T_R$. This temporal factor determines the characteristics of the recovery phase and has a greater impact on the calculated resilience than does the potential increase in redundancy. This is particularly true when the switching probability $p_s$ is low. An important long term challenge is to model adaptation, which, according to the National Academy of Sciences, is part of the response cycle that follows restoration and includes all activities that enable the system to better resist similar adverse events in the future.

Ganin et al. (2015) note that the main advantages of the approach include its applicability to any system that can be represented as a set of networks. Also both the resilience and the robustness of a system are metricized using a real value in range between 0 and 1 (where 1 corresponds to the perfect resilience or robustness) making comparison of resilience of different systems easy. On the other hand mapping the resilience property of a system to a single value necessarily shadows some system's important characteristics (for instance, the rate of recovery). The resilience profile could be used as a more holistic representation of the system's resilience noting that even in that case only the average value of critical functionality (at each time step) is taken into account. To fully describe a system one should consider the distribution of the value of critical functionality (at each time step) for different initial adverse events. Finally, it is not possible to simulate all adverse events from the range used to estimate resilience and the approach is Monte-Carlo based. It means that in order for the results to be reliable the number of simulations is typically required to be very high.

## *1.5* Tips for Generating Metrics

### 1.5.1 Generalized Metric Development Process

The following process towards the development of metrics is adapted by McKay et al. 2012.

1. Objective Setting: Articulate clear, specific goals. This should be done in a structured manner. Gregory and Keeney (2002) outline a structured approach to do this.
    a. Write down all of the concerns that the project team feels is relevant.
    b. Convert those concerns into succinct verb-object goals (e.g., minimize downtime).
    c. Next, these should be organized, often hierarchically, separating goals which represent means from those which represent ends.
    d. Finally, review and clarification should be conducted with the project team. This may be an iterative process.
2. Develop Metrics: Once the objectives are clearly articulated and organized, metrics can be formally developed.
    a. The first step is to select a broad set of metrics, which may be selected from existing lists or guidelines, or created by a project team or subject matter experts for the particular purpose at hand. This step is where the Resilience Matrix could facilitate metric development.
    b. Next, this set of metrics should be evaluated and screened to determine whether it meets the project objectives and the degree to which the metrics meet the desirable qualities of metrics, explained earlier in this chapter. At this stage, remaining metrics can be prioritized.

c. Finally the remaining metrics should be documented, including assumptions and limitations, and other supporting information.
3. Combination and Comparison: A method should be developed for how the metrics will ultimately be used to support decision making and drive action. Some methods include:
    a. Narrative Description: Simple techniques where trade-offs may be simple such as listing evidence or best professional judgement.
    b. Arithmetic Combination: Simple mathematical techniques for combining dissimilar metrics such as simple aggregation of metrics with similar units (e.g., cost), converting to similar units (e.g., monetization), or normalizing to a similar scale (e.g., 0 to 1).
    c. Multi-Criteria Decision Analysis: A method for weighting and scoring dissimilar decision criteria based on their relative importance and performance with respect to an objective.
    d. Interdependent Combination: For systems that are complex, usually involving intricate internal relationships, more intensive modeling efforts may be necessary, such as Bayesian networks or other complex systems modeling techniques.

The above-mentioned process, along with a solid metric development process, can greatly aid in devising effective metrics. Often it is necessary to develop a conceptual model of the system in order to identify the functional relationships and critical elements and processes within a system. This can be done using a Network Science approach described above.

### 1.5.2 Best Practices in Metric Development and Validation

Validation of metrics is an often overlooked aspect of the metric development process. Neely et al. (1997) provide some questions to ask regarding whether the output from the metrics is appropriate, specifically whether the metrics have a specific purpose, are based on an explicit formula and/or data source, and are objective and not based solely on opinion (Neely et al., 1997). Similarly, Eckerson (2009) lays out a series of questions that can serve as a quality check on developed metrics, to ensure that they are of high quality:

- Does it link to strategy?
- Can it be quantified?
- Does it drive the right behavior?
- Is it understandable?
- Is it actionable?
- Does the data exist?

Regarding the number of metrics necessary, it isn't necessarily the quantity of metrics that constitute a successful implementation, but whether these metrics are collectively comprehensive enough to address everything deemed important (McKay et al. 2012). Eckerson (2009) recommends that a set of metrics be *sparse*, since with a limited number of metrics it is easier to analyze how metric-level changes drive the performance in the system, as well as the practical fact that gathering, synthesizing, and presenting multiple data streams often takes quite some time. More granular, process-level metrics may still be required however, and Eckerson (2009) proposes a MAD (monitor, analyze, drill) framework for presenting different levels of resolution to different users of that information.

Another ongoing element of validation is traceability, as evidenced in the framework presented by Neely et al. (1997), which includes a list of information (known as the performance measure record sheet) such as how often data is to be collected, and by whom, as well as important questions such as "who acts on the data?" and "what do they do?". If these questions are considered and answered as the need arises, it is known who is responsible for making the measurement and what actions are to be taken as a result. This can reveal insight into the metric and how they are measured and being utilized, not just for the current project but for future reference. An item on the list asks what the metric "relates to." This can assist in entering the mindset of approaching metrics with an interconnected and goal-oriented viewpoint.

Other validation-related efforts include standardizing methods for ICS metric development and implementation, as well as institutionalizing a clear means to integrate metrics with decision analytic tools to support the risk management process. Finally, given the dynamic nature of cyber threats, periodic review and updating of ICS metrics should be conducted to keep abreast of the latest developments in the field.

## 1.6  Conclusions

Despite existing guidelines and frameworks, designing and managing for security in cyber-enabled systems remains difficult. This is in large part due to the challenges associated with the *measurement* of security.  A critical element in eliciting a meaningful metric is in gathering the relevant information about one's system and aligning that metric with measurable goals and strategic objectives.  For ICSs, time, safety and continuation of services factor considerably into overall goals, since many systems are in a position where a failure can result in a threat to human lives, environmental safety, or production output.  Often it is necessary to develop a conceptual model of the system or develop a standardized list of questions or topics helps to identify critical process elements, the functional relationships and critical elements and processes within a system. In this chapter, we discuss in detail two approaches for the generation of broadly applicable security and resilience metrics and their integration to quantify system resilience. The first method is a semi-quantitative approach in which the stages of the event management cycle (plan/prepare, absorb, recover, and adapt) are applied across four relevant domains (physical, information, cognitive, social), forming a matrix of potential security metrics. Second is a quantitative approach based on Network Science, in which features such as network topologies can be modeled to assess the magnitude and responsiveness of the critical functionalities of networked systems. Validation of metrics is an often overlooked aspect of the metric development process; however a series of questions can serve as a quality check on developed metrics, to ensure that they are of high quality.